\documentclass[12pt]{article}
\usepackage{amsmath}
\usepackage[dvips]{graphicx}
\usepackage{float}
\usepackage[centerlast,small]{caption}
\newcommand{\p}{\ensuremath{\mathbf{P}^1}}
\newcommand{\px}{\ensuremath{\mathbf{P}^1_x}}
\newcommand{\pz}{\ensuremath{\mathbf{P}^1_z}}
\newfloat{Figure}{h}{Figure }

\begin{document}
\title{Geodesics and BPS States in N=2 Supersymmetric QCD}
\author{Jeffrey M. Rabin \\ Dept. of Mathematics, U.C.S.D. \\ La
Jolla, CA 92093 \\ jrabin@ucsd.edu}
\maketitle
\begin{abstract}
Conformal mapping techniques are used to determine
analytically the geodesics on the Seiberg-Witten Riemann surface
which correspond to the BPS dyon states in $N=2$ SUSY QCD with
gauge group $SU(2)$ and
up to three flavors of massless fundamental matter.
The results are exact for zero and two flavors, and approximate in
the weak-coupling limit for one and three flavors.
The presence of states of magnetic charge 2, in the three-flavor case
only, is confirmed.
\end{abstract}

\section{Introduction}
The Seiberg-Witten (SW) solution of $N=2$ supersymmetric Yang-Mills
theory with gauge group $SU(2)$ and $N_f$ flavors of fundamental
quarks, $0 \le N_f \le 3$, is based on assumptions about the
spectrum of Bogomol'nyi-Prasad-Sommerfield (BPS) states in this theory, 
specifically the charges of
the particles which become massless at the singularities of the
moduli space \cite{sw1,sw2}.
These assumptions have been partially checked by several different
methods.
In the weak-coupling limit, semiclassical and adiabatic arguments
lead to a description of the dynamics of slow, heavy dyons in terms
of supersymmetric quantum mechanics on the moduli space $M_g$ of
monopoles of magnetic charge $g$ \cite{gauntlett94}.
$L^2$ index theory on $M_g$ provides information about the spectrum
of this theory \cite{sethi,gauntlett}.
In particular, the existence of dyons of charge $g=2$ in the
case $N_f=3$ is demonstrated.
However, at present these methods require explicit knowledge of the
metric on $M_g$ and the connection on the index bundle of the Dirac
operator over $M_g$.
Therefore they are currently limited to charges $g \le 2$.
Arguments based on monodromy at the singularities and the existence
of a curve of marginal stability separating weak and strong coupling
regions in the moduli space \cite{bilal1,bilal2} 
confirm that the SW 
assumptions are the only consistent ones, but do not directly
demonstrate the existence of the assumed dyon states. 

Recently, another approach has been developed based on the
derivation of the SW description from the low-energy
limit of type IIB string theory \cite{kachru,klemm,e6,brandhuber}.
The ten-dimensional string theory is compactified on a Calabi-Yau
manifold which is a $K3$ fibration over \p.
This \p\ with its standard complex coordinate $z$ will be denoted
\pz.
The relevant BPS states in this theory are associated with D-branes
wrapped around nontrivial 3-cycles of the Calabi-Yau manifold.
Their masses are given by the volumes of the cycles as measured by the
holomorphic 3-form. 
At finitely many ``branch points" in \pz\ the $K3$ fibers degenerate by the
vanishing of a 2-sphere.
The relevant 3-cycles are either $S^2 \times S^1$, where the
first factor is in the fiber and the second in the base, or $S^3$.
The latter can be viewed as a singular case of $S^2 \times I$, where
the endpoints of the interval $I$ in the base are two of the branch
points at which the $S^2$ vanishes.
In the low-energy limit this information about the homology is
captured in the SW Riemann surface, which is also a
``fibration" (branched cover) of \pz.
The $K3$ fibers are replaced by discrete sets of points which may be
thought of as two endpoints of a diameter of each nontrivial $S^2$
in the $K3$.
Pairs of these points coincide, and sheets of the Riemann surface
cross, over the same branch points on \pz\ where an $S^2$
vanished.
BPS saturated states are D-branes wrapped around minimal volume
cycles in their homology classes.
This requires that the $S^2$ be minimal, and then that the $S^1$
or $I$ be (the image in \pz\ of) a closed or open geodesic in the
SW Riemann surface for a metric induced by the holomorphic 3-form.
This (flat) metric turns out to be $|\lambda(z)|^2$, with 
$\lambda(z)$ the meromorphic differential 1-form of the SW theory.
The homology class of the geodesic determines the electric and
magnetic charges of the BPS dyon state in the supersymmetric gauge
theory.
(Strictly speaking, an open curve defines a homology class
only relative to its endpoints;
reference to the Calabi-Yau setting shows that the charges can
be defined by the intersection numbers of the geodesic with a
homology basis chosen to avoid the endpoints.)

These geodesics have been computed by numerical integration of the
geodesic differential equation for $N_f=0$ \cite{klemm}, $N_f=1,2,3$
\cite{brandhuber}, and in some examples from F-theory
\cite{johansen}.
In most cases the results confirm the expected BPS spectrum.
However, the expected $g=2$ states for $N_f=3$ were not found in
\cite{brandhuber}.
An interesting feature of the method is its sensitivity to the
1-form $\lambda$ itself, not just its periods.
We assume in this paper that the 1-form suggested by the integrable
systems approach to SW theory \cite{martinec,itoyama,gorsky} 
is the correct one.
This has been partially confirmed by deriving it from string theory for
$N_f=0$ \cite{klemm} and $N_f=1$ \cite{nam}.

In this paper we will describe an analytic method of determining the
geodesics based on the conformal mapping to the flat coordinates
for the metric, which is just a classical Schwarz-Christoffel
mapping.
The geodesics of course map to straight lines.
For $N_f=0,2$ they can be determined exactly at both weak and strong
coupling, and the transition can be understood as a change in
convexity of the image of the mapping.
For $N_f=1,3$, they can be described to a good approximation in the
weak-coupling limit.
The qualitative behavior of the geodesics is easily understood in
this framework, including the typical spiral behavior around branch
points.
The presence of $g=2$ states for $N_f=3$ is also clear despite the
approximate treatment, although their absence for $N_f=1$ is less
so.
The distinction between even and odd $N_f$ is due to the symmetry of the
SW curve $F(x,z)=0$ under $x \rightarrow -x$ in the
even case.
Our discussion is restricted to the case of massless matter.

The SW curve can be viewed as a branched cover of either
the $z$ plane (\pz) or the $x$ plane (\px).
Because the form of $\lambda$ is simpler and more uniform in $N_f$
when expressed in terms of $x$, we will generally work in 
\px .
However, it is the $z$ plane which plays the distinguished role as
the base of the $K3$ fibration in the derivation from string theory.
Thus, the geodesics should join the branch points in \pz,
even if they are viewed in projection on \px.
(This probably explains the failure of \cite{brandhuber} to find the
$g=2$ states.)
For $N_f=0$ we will be able to describe the relation between the
projections of the geodesics onto the two planes quite precisely.

In Section 2 we review some conformal mapping techniques, namely the
Schwarz-Christoffel mapping and the Schwarz reflection principle.
Section 3 recaps the SW curves for various $N_f$,
their branch points in both the $x$ and $z$ planes, and the
differentials $\lambda$. 

Section 4 treats the cases of even $N_f$.
The Schwarz-Christoffel mapping gives a simple and exact description
of the image of the $x$ plane for real values of the moduli space
coordinate $u$.
For $N_f=0$ we describe the geodesics in both weak- and
strong-coupling limits, and the transition between these.
The explicit mapping between the $x$ and $z$ planes is also
described.
$N_f=2$ exhibits the new feature arising from the presence of
matter: the opening of a ``window" between the right and left half
$x$ planes.
Because of the positioning of the branch points, it is not exploited
in this case, however.
Still, we make use of the exact solution of this case to draw
lessons about the behavior of the geodesics in such regions which
can be applied to the remaining cases with matter.

Section 5 discusses the cases $N_f=1,3$.
These differ in the positioning of the branch points in relation to
the window.
Unfortunately we cannot identify any convenient region of the
left $x$ plane whose image is bounded by straight line segments to
which the Schwarz reflection principle applies.
Therefore our treatment of the geodesics is less complete.
However, we can say enough about their behavior to argue for
the presence of the $g=2$ dyon states for $N_f=3$.

Section 6 contains some concluding remarks.

As this paper was being completed, additional works appeared in which
analytic methods for determining the geodesics were proposed
\cite{fayyazuddin,schulze}.

\section{Geodesics and Conformal Mapping}

The metric $|\lambda(x)|^2$ induced by the SW differential on \px\ 
is flat; one can go to flat coordinates $w$ by the conformal mapping
$dw = \lambda(x)$ or
\begin{equation}
w = \int^x \lambda(x).
\end{equation}
In all cases of interest here $\lambda$ has the form
\begin{equation}
\lambda(x) = \prod_i (x-x_i)^{\alpha_i} dx,
\end{equation}
exhibiting zeros and branch points at the various $x_i$.
Such a conformal map is an example of a Schwarz-Christoffel mapping
\cite{henrici}.
In our cases we will always have $\sum_i \alpha_i = 0$, so that the
map approaches the identity for large $x$.
The image of a straight line under such a map can be determined by 
writing
\begin{equation}
\arg dw = \arg dx + \sum_i \alpha_i \arg (x-x_i).
\end{equation}
As $x$ moves along the line, $w$ moves along a curve whose direction
is locally given by $\arg dw$, hence determined by the variation of
the $\arg (x-x_i)$.
If, in particular, the $x_i$ are all real and we ask for the image
of the real axis, then $\arg dw$ is piecewise constant, changing by 
$-\pi \alpha_i$ as $x$ passes $x_i$ from left to right.
The image consists of straight line segments joining the images of
the $x_i$.
The image of the upper half plane lies to the left of this 
polygonal path.
We will apply these ideas to slightly more general regions, namely 
quadrants.

We recall also the Schwarz reflection principle \cite{henrici}.
This describes the analytic continuation of a conformal mapping
across a straight line segment in the domain whose image is also
straight.
In the above context, if $x$ passes from the upper to the lower half
plane across a segment of the real axis, the image point passes
from the image of the upper half plane into a region which is its
reflection across the particular boundary segment which is crossed.
The distinct regions resulting from reflection across different
segments display the monodromy properties of the mapping around the
branch points.

\section{The Curves}

The hyperelliptic curves describing $N=2$ SUSY $SU(2)$ QCD with $N_f$
flavors of massless matter are given by \cite{martinec,hanany}
\begin{equation}
z + \frac{Q(x)}{z} = 2P(x),
\end{equation}
where $Q(x) = \Lambda^{4-N_f}x^{N_f}$ and
\begin{alignat}{2}
P(x) &= x^2 - u, & \qquad N_f &= 0,1. \notag \\
P(x) &= x^2 - u + \frac{\Lambda^2}{8}, & \qquad N_f &= 2. \notag \\
P(x) &= x^2 - u + \frac{\Lambda}{4} x, & \qquad N_f &= 3.
\end{alignat}
$\Lambda$ is the dynamically generated scale and the moduli space
is the complex $u$ plane.
In the weak-coupling limit $u \rightarrow \infty$ which will be our
main concern we can take $P(x)=x^2-u$ in all cases.
We will always consider real positive values of $u$.
We will also set $\Lambda = 1$.
The change of variables 
\begin{equation}
y = z - P(x) = \frac{1}{2} [z - \frac{Q(x)}{z}]
\end{equation}
brings the curves to the standard hyperelliptic form 
\begin{equation}
y^2 = P(x)^2 - Q(x).
\end{equation}

The polynomial $P(x)^2 = (x^2-u)^2$ has, of course, two double zeros at 
$x=\pm \sqrt{u}$.
For $N_f$ even, the perturbation $-Q(x) = -x^{N_f}$ is negative at
each zero and therefore splits the double roots into two pairs of
real roots given for large $u$ by
\begin{equation}
\pm u^{1/2} \pm \frac{1}{2} u^{(N_f-2)/4},
\end{equation}
with the two signs chosen independently.
These two pairs of branch points in the $x$ plane are related by the
symmetry under $x \rightarrow -x$.
This symmetry is absent for $N_f$ odd, and indeed the perturbation
is then positive at $x=-\sqrt{u}$, which therefore splits into a
pair of {\it complex} branch points,
\begin{equation}
- u^{1/2} \pm \frac{i}{2} u^{(N_f-2)/4},
\end{equation}
while the positive root still splits as above.

Within the framework of supersymmetric gauge theory, the
Seiberg-Witten differential $\lambda$ is not determined within its
cohomology class.
However, both the integrable systems approach \cite{martinec,gorsky}
and the derivations via the low-energy limit of string theory
\cite{klemm,nam} suggest the specific representative
\begin{equation}
\lambda = x \, d \log \frac{z}{\sqrt{Q(x)}} = 
\frac{dx}{y} \frac{2QP' - PQ'}{2Q}.
\end{equation}
For our curves in the weak-coupling limit we find
\begin{equation}
\lambda = \frac{x \, dz}{z} - \frac{N_f}{2} dx =
\frac{dx}{2y}[(4-N_f)x^2 + N_f u],
\end{equation}
although we will choose the normalization
\begin{equation}
\lambda = \frac{dx}{y}\left(x^2 + \frac{N_f}{4-N_f}u \right).
\end{equation}
Note the pair of zeros on the imaginary axis at 
$x=\pm i \sqrt{\frac{N_f}{4-N_f}u}$ in addition to the branch points
at the above zeros of $y$.

Generically, a distance measured by the metric $\lambda$ is of
the same order as the coordinate distance in the $x$ plane, the
basic scale being set by $\sqrt{u}$.
However, distances are increased near branch points of $\lambda$,
e.g. by $\log u$ factors, and decreased near its zeros.

We will also need the (finite) branch points in \pz, the endpoints of the
relevant geodesics.
These are the points $z$ at which 
\begin{equation}
z + \frac{x^{N_f}}{z} = 2(x^2-u)
\end{equation}
has fewer than the generic number of solutions for $x$.
For $N_f < 3$ this generic number is two, and by the $x \rightarrow
-x$ symmetry the branch points must occur for $x=0$ when $N_f$ is even.
For $N_f=0$ we find
\begin{equation}
z = e_\pm = -u \pm \sqrt{u^2-1},
\end{equation}
or, at weak coupling, $e_- = -2u$ and $e_+ = -1/2u$.
For $N_f=2$ we have simply $e_- = -2u, \; e_+ = 0$.

For $N_f=1$ the solution for $x$ is
\begin{equation}
x = \frac{1}{4}\left[ z^{-1} \pm \sqrt{z^{-2} - 8(z+2u)}\right].
\end{equation}
The branch points are the roots of a cubic equation and for large
$u$ are given by
\begin{alignat}{2}
z &= -2u & \qquad (x &= -1/8u), \notag \\ 
z &= \pm \frac{1}{4\sqrt{u}} - \frac{1}{64u^2} &
\qquad (x &= \pm \sqrt{u} + \frac{1}{16u}).
\end{alignat}

For $N_f=3$ the generic number of solutions for $x$ is three, and
there are branch points at which generically pairs of these coincide.
They are the zeros of the discriminant $\Delta$ of the cubic
\begin{equation}
x^3 - 2zx^2 + z(z+2u) = 0,
\end{equation}
namely
\begin{equation}
\Delta = \frac{1}{4} z^2 (z+2u) 
\left[(z+2u) - \frac{32}{27}z^2 \right].
\end{equation}
The branch points are at $z=0$, corresponding to $x=0$; $z=-2u$,
corresponding to $x=-4u$ and $x=0$; and the roots of the quadratic,
which are approximately
\begin{equation}
z = \pm \frac{3\sqrt{3}}{4} \sqrt{u} \;\;\;\;\; 
( x = \pm \sqrt{3u}, \; \mp \frac{\sqrt{3}}{2} \sqrt{u} ).
\end{equation}

\section{The Cases $N_f = 0,2$}

The SW curve for $N_f=0$ is $z + z^{-1} = 2(x^2-u)$, or 
$y^2 = (x^2-u)^2 - 1$, and the conformal mapping which flattens the
metric on the $x$ plane is
\begin{equation}
w = \int^x \frac{x^2 \,dx}{\sqrt{(x^2-u)^2-1}}.
\end{equation}
The double zero of $\lambda(x)$ at the origin causes the map to locally triple
angles there, so the map is not one-to-one on the upper half-plane.
Therefore it is convenient to study the image of a quadrant instead.
Figure 1 shows the image of the first quadrant and some of its
Schwarz reflections representing the fourth quadrant for 
weak coupling, $u >> 1$.
(All figures in this paper show qualitative behavior correctly, 
but are not necessarily strictly to scale.)
The image of the imaginary axis is a straight line thanks to the 
symmetric placement of the branch points, which causes the changes
in the $\arg (x-x_i)$ to cancel as $x$ moves down this axis.
Reflecting across the imaginary axis, we see that the images of the
first and second quadrants overlap in a rectangle, which is the
region covered twice in mapping the entire upper half-plane.
The sides of this rectangle are the SW half-periods, 
$\frac{\pi}{2} \sqrt{u}$ (vertical) and $\sqrt{u} \log u$
(horizontal).
The geodesics we seek are straight lines from $x=0$ (the image of
both branch points in \pz) to one of its reflected images.
Such geodesics can meet the branch cut joining the branch points on
the positive real axis at most once (at an endpoint), 
corresponding to the magnetic charge, but by passing through many
reflected regions may cross the rest of the positive real axis an
arbitrary number of times, giving arbitrary electric charge.
In the $x$ plane, these geodesics are somewhat singular: they leave
$x=0$, spiral around the cut until they reach one of its endpoints,
and then retrace the same path backwards to $x=0$.
By the $x \rightarrow -x$ symmetry there are similar geodesics in
the left half-plane, but none which cross the imaginary axis.
A small change in the initial direction produces a geodesic which
loops around the endpoint of the cut rather than terminating there,
follows almost the same path backwards, turns around just short of
$x=0$ and runs out to infinity.

\begin{Figure}
\begin{center}
\includegraphics[width=10cm]{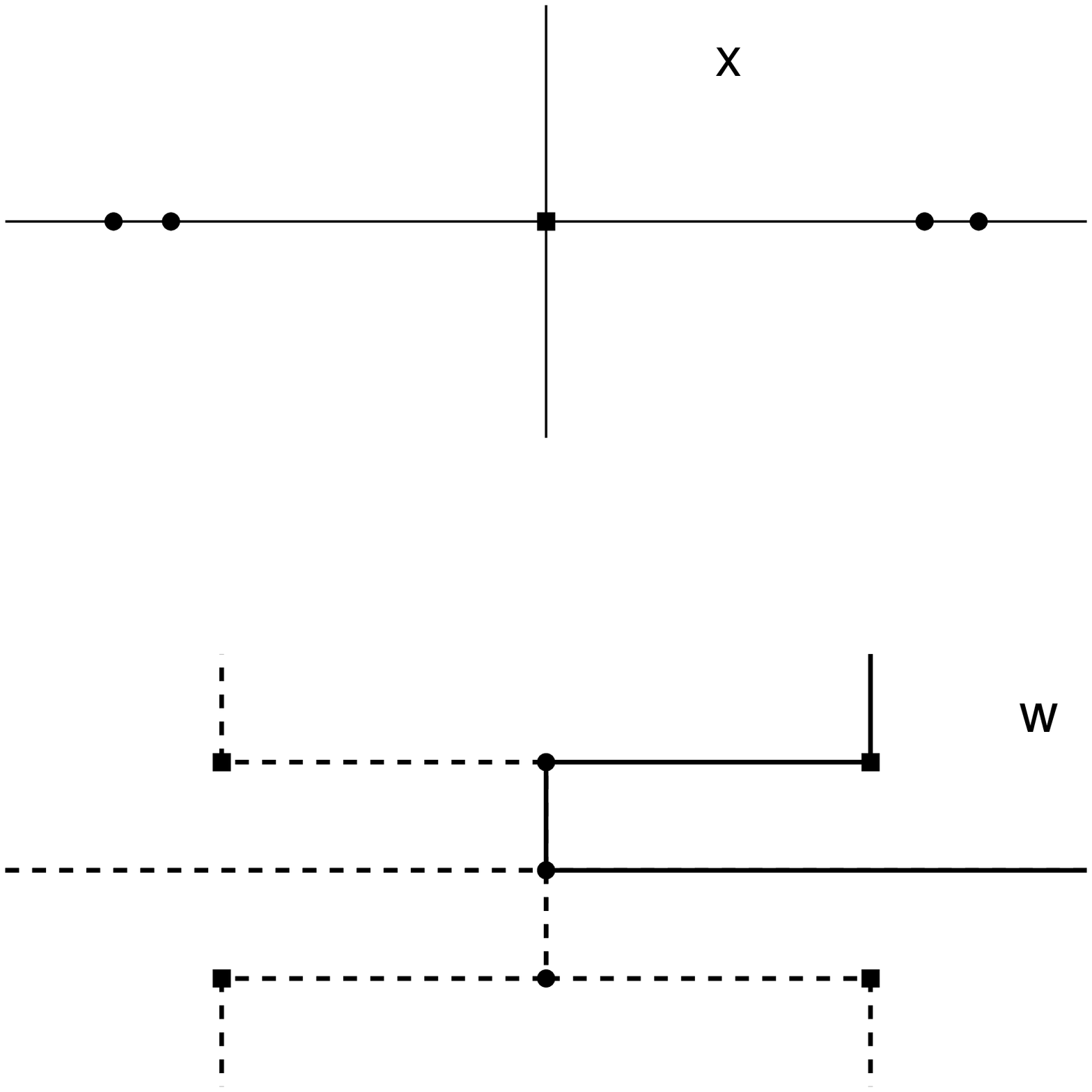}
\end{center}
\caption{The image of the first quadrant in the $x$ plane
lies to the right of the solid boundary in the $w$ plane. Branch
points on \px\ are labelled by dots, those on \pz\ by squares.
Reflections of this region across segments of the positive real axis
are also shown. Straight lines joining the squares acrue one unit of
magnetic (electric) charge in crossing a vertical (horizontal)
boundary between regions.}
\end{Figure}

Due to the simplicity of this case we can trace the geodesics back
to the ``physical" $z$ plane.
Introduce an intermediate $s$ plane by
\begin{equation}
z + \frac{1}{z} = 2s = 2(x^2-u).
\end{equation}
The map from $x$ to $s$ doubles angles to expand the right $x$ plane
to the entire $s$ plane, sending the origin to the point $s=-u$
and the branch cut to $[-1,1]$.
The map from $z$ to $s$ is a standard Joukowski transformation
sending the exterior as well as the interior of the unit circle to
the cut $s$ plane. 
The branch points $e_-,e_+$ in the $z$ plane lie in the exterior and
interior of the circle, respectively.
The $g=1$ geodesics which met the cut now cross this circle and
indeed join $e_-$ with $e_+$, while the $g=0$ geodesics are now
closed curves with both endpoints at, say, $e_-$, 
as expected for the gauge bosons.
These mappings are illustrated in Figure 2.

\begin{Figure}
\begin{center}
\includegraphics[width=10cm]{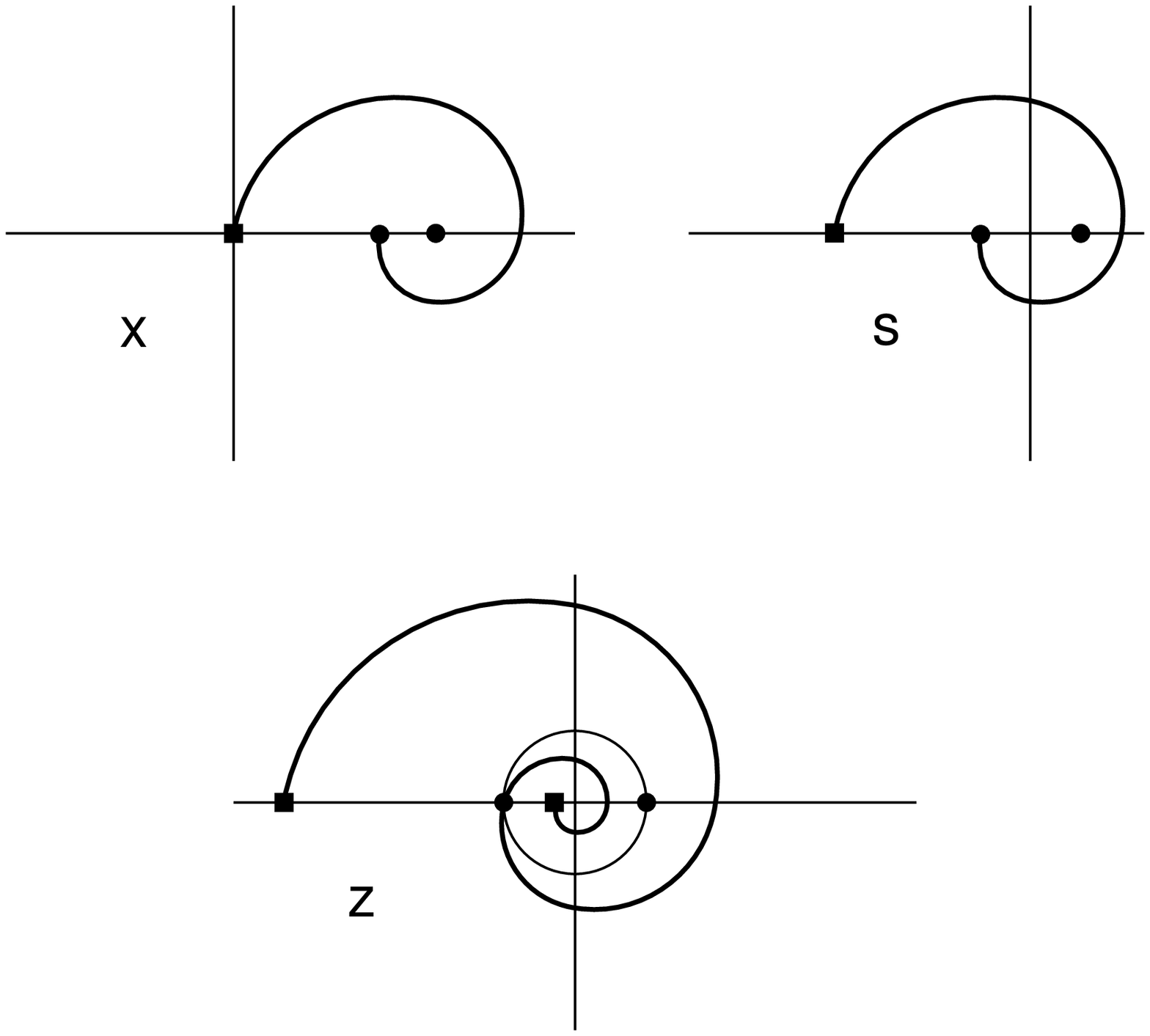}
\end{center}
\caption{Views of a $(g,q)=(1,2)$ geodesic in the $x$,
$s$, and $z$ planes.}
\end{Figure}

We can also study the conformal mapping of the $x$ plane in the
strong-coupling case $0<u<1$ and understand the transition as $u$
crosses $x=1$, where the curve of marginal stability meets the real
axis.
As $u$ decreases, the branch points on \px\ move toward the origin.
When $u=1$ the inner pair of branch points meet there, and the
homology cycle encircling this pair vanishes.
As $u$ decreases further these points separate along the imaginary
axis, leaving one branch point on each half axis at strong coupling.
In this process the homology cycle dual to the vanishing cycle is
shifted by half its Picard-Lefschetz monodromy around the
singularity.
Figure 3 shows the image of the first quadrant in this case.
Due to the change in convexity of the image region, the only
straight lines from $x=0$ to an image point which remain in the
region are now the vertical and horizontal lines.
These would represent pure electric or magnetic charges if not for
the change in the homology basis which occurs in passing the
singular point $u=1$ \cite{bilal1}.
In fact they correspond to the weak coupling states
$(g,q)=(0,1),(1,1)$.
We have thus understood completely the representation of BPS states
by geodesics in this case.

\begin{Figure}
\begin{center}
\includegraphics[width=12cm]{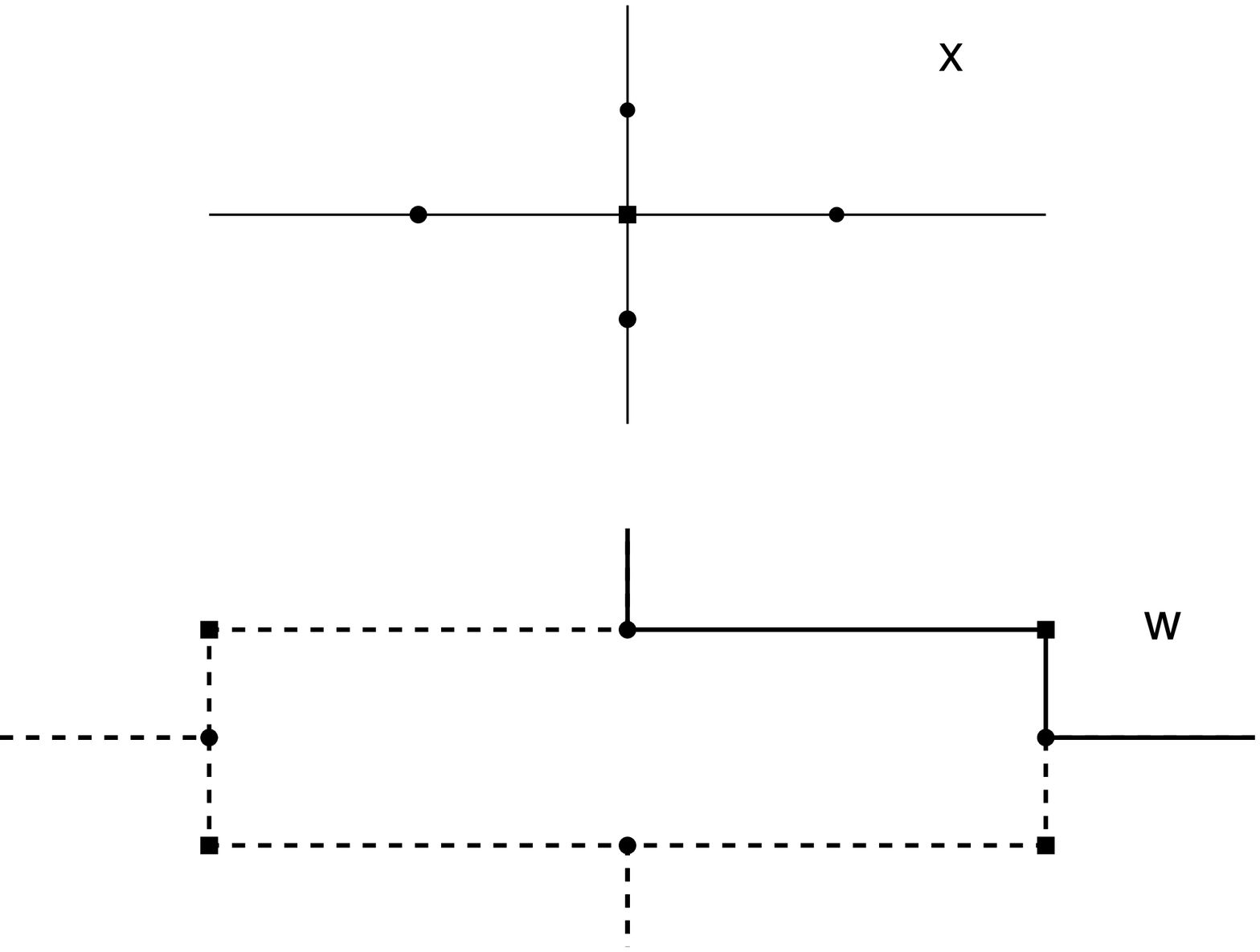}
\end{center}
\caption{For strong coupling the image of the first
quadrant together with some of its reflections forms the exterior of
the rectangular region shown. Straight lines joining the branch
point images cannot enter the rectangle and must lie on its
boundary.}
\end{Figure}

The two-flavor case is very similar.
We treat only the weak-coupling limit.
The curve still has the $x \rightarrow -x$ symmetry which leads to a
symmetric placement of branch points in \px\ and puts both of the
\pz\ branch points at $x=0$.
The major change in the mapping,
\begin{equation}
w = \int^x \frac{(x^2+u)\, dx}{\sqrt{(x^2-u)^2 - x^2}},
\end{equation}
is that the double zero of $\lambda(x)$ 
at the origin has split into two zeros on
the imaginary axis.
Figure 4 shows the resulting image of the first quadrant.
The part of the imaginary axis between the origin and the new zero
is mapped to a small ``window" which can be crossed from left
to right, leading by reflection into an image of the second 
quadrant.
However, the geodesics which start and end at $x=0$ cannot take
advantage of this possibility, and behave just as for
$N_f=0$, as expected.
(A geodesic which starts at $x=0$ and passes through the window may
spiral around $x=0$ but will eventually run out to infinity.)
The existence of the window will persist in the odd $N_f$ cases
which are not exactly solvable by our methods.
Its size is $O(\sqrt{u})$ but the numerical coefficient grows with
$N_f$.
(Its size relative to the vertical segment in the $w$ plane 
which is the image of the
branch cut is $6\%$, $18\%$, and $39\%$ in the cases $N_f=1,2,3$
respectively.)
Therefore, let us study the qualitative behavior of other geodesics
in the present case, to develop intuition
which will apply more generally.
A geodesic which crosses the cut at a very small angle will then pass
through many reflected regions in the vertical direction, meaning
that it spirals outward around the cut many times.
If it misses the window it will run out to infinity.
If it passes through the window it enters the left half-plane and
spirals inward toward the cut there, eventually crossing that cut.
It then begins spiraling outward and the behavior repeats.
Very small changes in the initial direction of the geodesic can make
it pass through any desired point on the real axis, and hit or miss
the window.
Increasing the angle at which the geodesic crosses the cut
decreases the number of spirals and thus the electric charge.
At weak coupling the cuts are well separated from each other and
from other features such as zeros, so the local spiraling behavior
around the cuts should occur in all cases independent of other
global differences. 

\begin{Figure}
\begin{center}
\includegraphics[width=10cm]{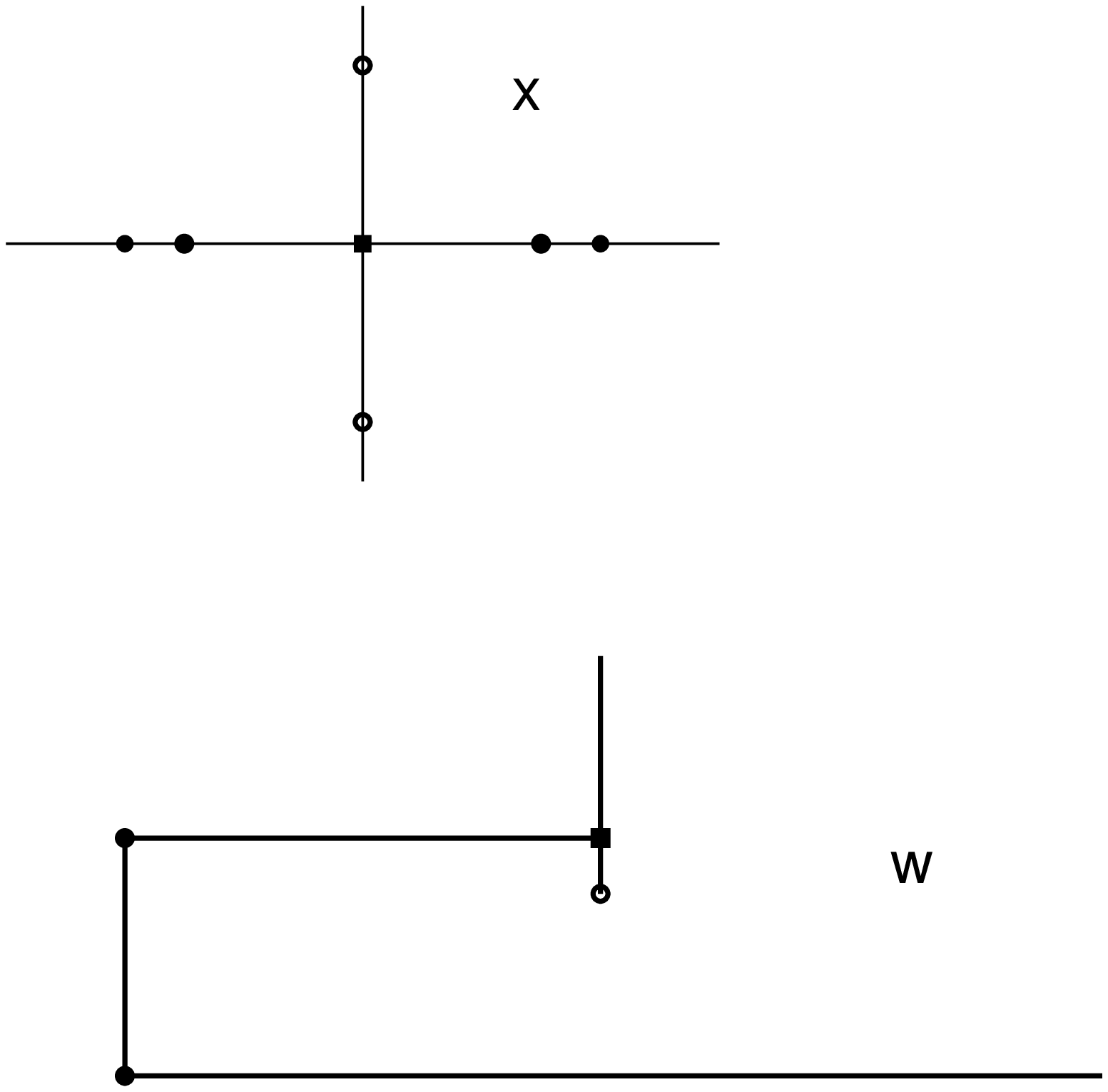}
\end{center}
\caption{The image of the first quadrant for $N_f=2$.
Open circles mark zeros of the SW differential
$\lambda$, which produce a small ``window'' in the image.
Straight lines joining the squares to their reflected images
do not pass through this
window but duplicate the behavior for $N_f=0$.}
\end{Figure}

\section{The Cases $N_f = 1,3$}
When $N_f$ is odd we lose the $x \rightarrow -x$ symmetry.
The branch points in \px\ form a real pair near $\sqrt{u}$ and a
complex conjugate pair near $-\sqrt{u}$.
At weak coupling the separation between the pairs is large while the
splitting within each pair is small (absolutely so for $N_f=1$,
relatively so for $N_f=3$).
The image of the imaginary axis under the conformal mapping is not
exactly straight due to the imperfect cancellation between the
$\arg(x-x_i)$ contributions of the pairs of branch points, but this
effect is small.
More consequentially, the image of the line $L$ through the pair of complex
branch points is not straight due to the $\arg(x-x_i)$ contributions
from the zeros on the imaginary axis and the distant pair of branch
points.
Figure 5 shows the images of the first quadrant and a region in the
second quadrant bounded by the real and imaginary axes and $L$.
The important feature is the placement of the branch points from
\pz\ in the image.
One lies on the branch cut near $\sqrt{u}$, one very near 
($O(u^{-1})$) the origin, and one very near ($O(u^{-1/4}$)) the
center of the cut near $-\sqrt{u}$.
Although we cannot use Schwarz reflection for the lines that are not
straight, it should be approximately valid where they are nearly
so. 
Further, the local spiral behavior of the geodesics near the branch
points is universal.
Geodesics leaving the ``positive" branch point cross the homology
generator encircling the cut, which contributes $1$ to the magnetic
charge. 
To obtain a state with magnetic charge 2, such a geodesic would have to
pass through the window, cross the cut near $-\sqrt{u}$, and end at
the ``negative" branch point.
The positioning of the negative branch point very near the cut
makes it quite hard to hit.
It requires that the image of the geodesic be a nearly
vertical line which passes through many reflected regions before
crossing the window.
However, it then enters the second (or third) quadrant 
and crosses the line $L$ far from the branch points, where the
reflection principle does not apply.
We conclude cautiously that there is no evidence for states of
magnetic charge 2 in the one-flavor theory, although charges 0 and 1
exist.

\begin{Figure}
\begin{center}
\includegraphics[width=12cm]{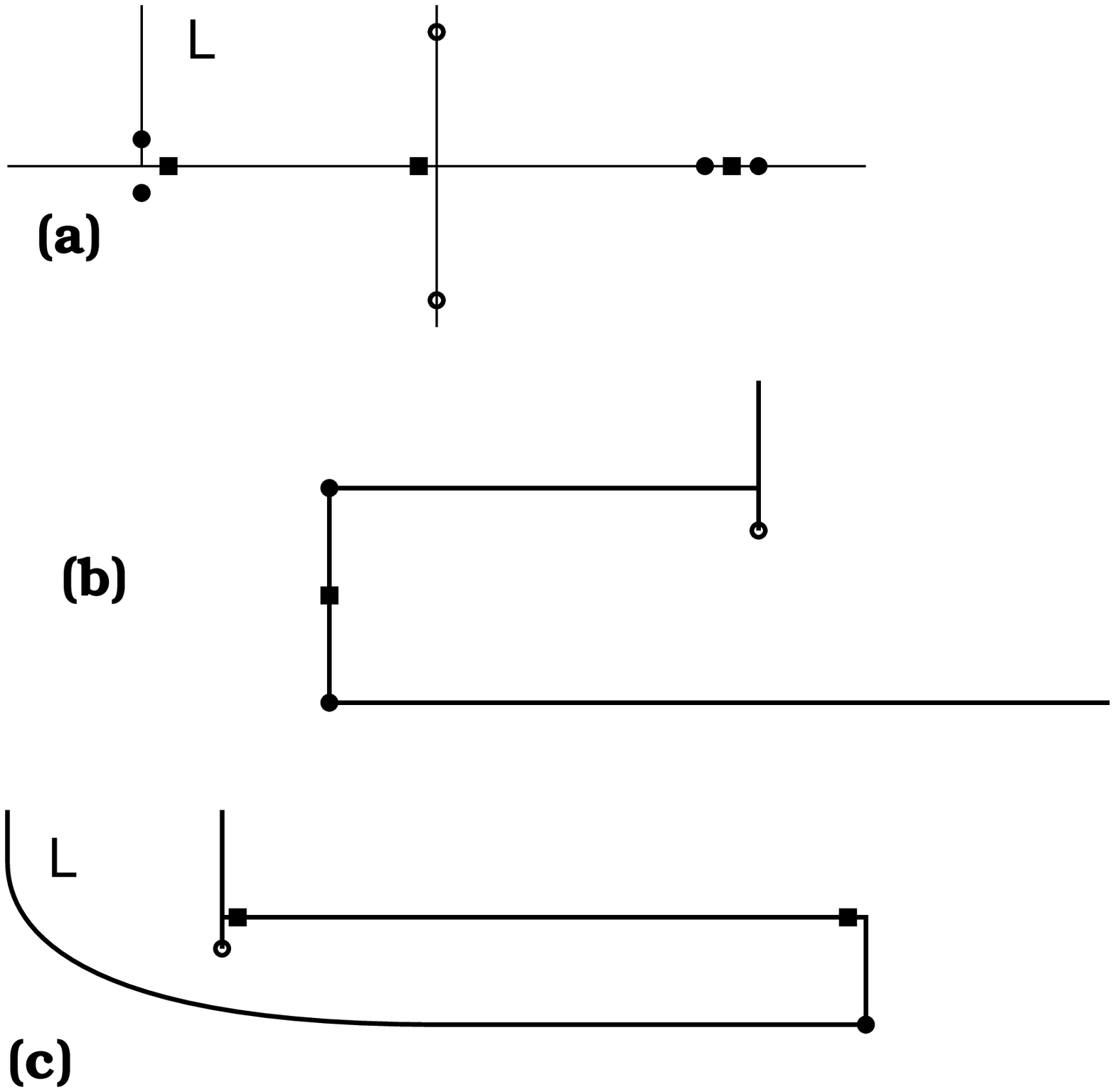}
\end{center}
\caption{The case $N_f=1$. 
The $x$ plane (a) is shown,
along with the images of the first quadrant (b) and the
semi-infinite rectangle in the second quadrant (c). 
A line entering (c) through the window, crossing the distant cut,
and ending on the reflection of the nearby square across it
would have to have slope of order $u^{3/4}$, and would have
met the curved portion of $L$.}
\end{Figure}

The placement of the branch points for $N_f=3$ is quite different,
as shown in Figure 6.
Most branch points have two images in \px\ , and their locations are
more generic in that they are not unnaturally close to branch cuts.
This allows them to be connected by lines whose slopes are not
unnaturally large. 
In addition to a variety of $g=0,1$ states, there are now geodesics
which leave one of the branch points on the positive real axis,
cross the cut there, pass through the (larger) window, cross the cut
in the left half-plane and spiral around it until they end on a
branch point there, crossing boundaries only where they are straight to a
good approximation.
Therefore the existence of these geodesics seems reliable despite
the approximation.
In terms of the qualitative behavior of geodesics gleaned from the
$N_f=2$ case, given that geodesics exist which cross both cuts,
small perturbations suffice to make them pass through arbitrary
points not too far away, including the desired branch endpoints.

\begin{Figure}
\begin{center}
\includegraphics[width=12cm]{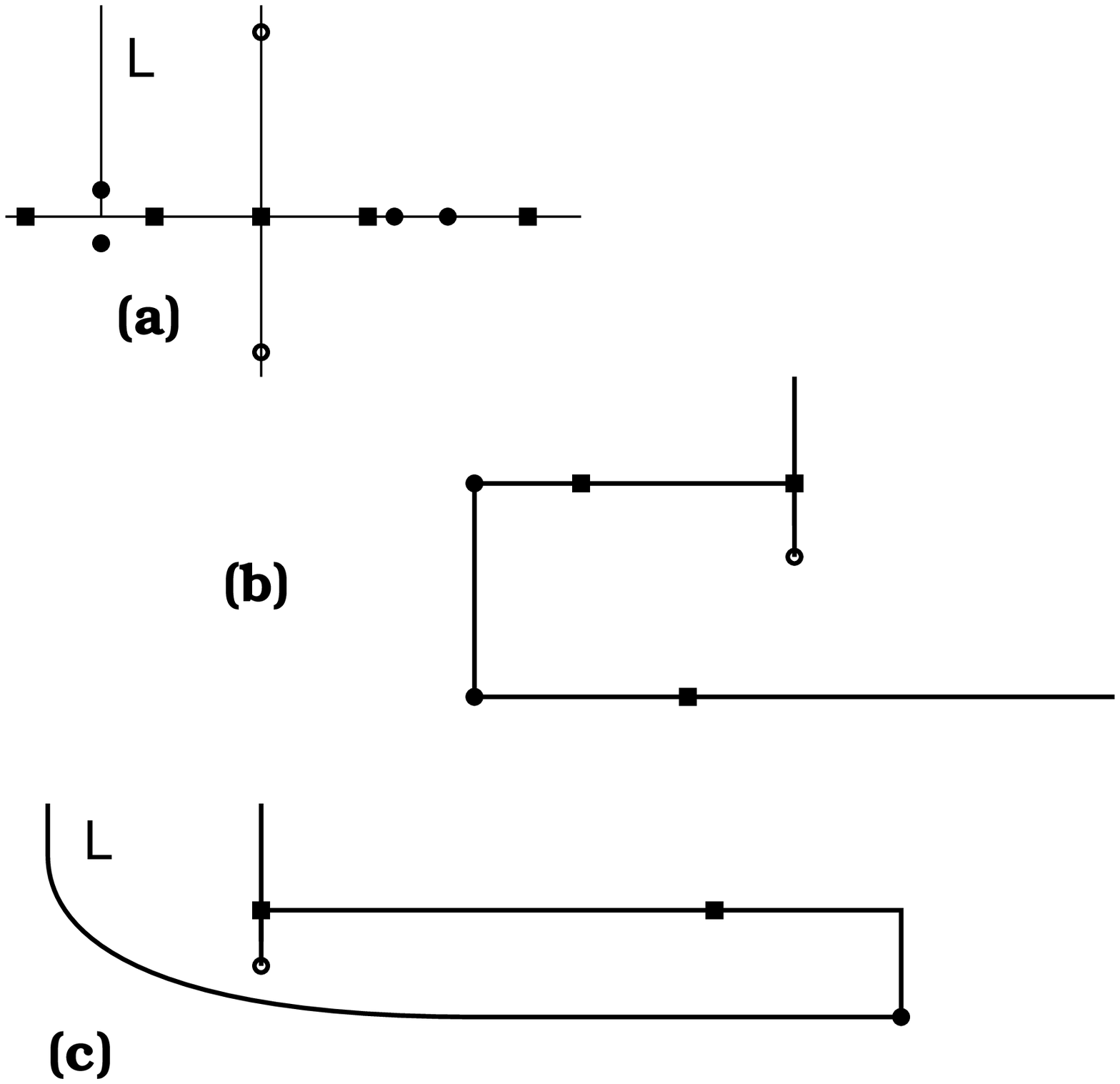}
\end{center}
\caption{The case $N_f=3$. The $x$ plane (a) and images of
the first quadrant (b) and the semi-infinite rectangle in the second
quadrant (c). In contrast to $N_f=1$, lines having slope of order
$1$ can cross the cuts in both quadrants to join branch points.}
\end{Figure}

\section{Conclusions}
We have used simple conformal mapping techniques to determine the
geodesics corresponding to the BPS states of $N=2$ supersymmetric
QCD with gauge group $SU(2)$, exactly for even $N_f$ and
approximately at weak coupling for odd $N_f$.
Qualitative features of the geodesics such as the typical spiral
behavior around branch cuts become transparent in this approach.
The important changes as $N_f$ is increased involve the placement 
of the branch points on \pz\ relative to those on \px.
The transition between weak and strong coupling involves a change in
convexity of the image region, which obviously alters the set 
of straight lines which can connect boundary points.
Although we considered only real positive values of $u$, for which the
classical Schwarz-Christoffel mapping applies, even when $u$ is
complex the convexity of the image region can only change
upon crossing the curve of marginal stability.
Indeed, the arguments of \cite{bilal1,bilal2}
concerning the spectrum outside and inside
this curve could be directly rephrased as arguments about the
behavior of the geodesics in these regions.

In our treatment it is quite clear that for $N_f=0,2$ the maximum
magnetic charge of an elementary BPS state is 1.
For $N_f$ odd the maximum charge is 2.
Such states seem to exist for $N_f=3$ due to the ``generic"
placement of the branch points, which allows them to be connected by
straight lines without unnatural ``fine-tuning".
For $N_f=1$, fine-tuning would be necessary to obtain such states
but our approximations are not reliable for such fine-tuned lines.
No such states are found in numerical studies \cite{brandhuber}.

These methods should be applicable to $N=2$ theories with other gauge
groups, possibly with massive matter.
The main requirement is that there exist a region of the moduli
space in which all branch points on \px\ are real, or nearly so.
Symmetries relating the branch points are also very useful.
One may hope that with analytic understanding of the geodesic
description of BPS states will come an understanding of the relation
between this description and that of $L^2$ index theory.

\section*{Acknowledgements}
I have enjoyed discussions with Albrecht Klemm, Harold Stark, 
Nicholas Warner and Eric Zaslow.

\end{document}